\documentclass[10pt,twocolumn, floatfix, altaffilletter, superscriptaddress,tightenlines, showpacs, showkeys, preprintnumbers, nofootinbib]{revtex4-1}
\pdfoutput=1
\usepackage[colorlinks=true,citecolor=blue,linkcolor=blue,breaklinks=true]{hyperref}
\usepackage{amsmath,amssymb}
\usepackage{epsfig} 
\usepackage{graphicx}        
\usepackage{url}
\usepackage{xcolor}
\usepackage{multirow}
\usepackage{placeins}
\usepackage{braket}
\usepackage{float}
\hypersetup{
  colorlinks=true,
  linkcolor=red,
  filecolor=magenta,   
  urlcolor=blue,
  citecolor=blue,
}

\allowdisplaybreaks

\bibpunct{[}{]}{,}{n}{}{,}

\def\beq{\begin{equation}}
\def\eeq{\end{equation}}
\def\bea{\begin{eqnarray}}
\def\eea{\end{eqnarray}}
\makeatletter
\@addtoreset{equation}{section}

\begin{document}

\title{Bremsstrahlung High-frequency Gravitational Wave Signatures \\ of High-scale Non-thermal Leptogenesis 
}
\author{Anish Ghoshal}
\email{anish.ghoshal@fuw.edu.pl}
\affiliation{Institute of Theoretical Physics, Faculty of Physics, University of Warsaw,
ul. Pasteura 5, 02-093 Warsaw, Poland}
\author{Rome Samanta}
\email{romesamanta@gmail.com}
\affiliation{CEICO, Institute of Physics of the Czech Academy of Sciences,
Na Slovance 1999/2, 182 21 Prague 8, Czech Republic}

\author{Graham White}
\email{graham.white@ipmu.jp}
\affiliation{Kavli IPMU (WPI), UTIAS, The University of Tokyo, Kashiwa, Chiba 277-8583, Japan}

\begin{abstract} 

Inflaton seeds non-thermal leptogenesis by pair producing right-handed neutrinos in the seesaw model. We show that the inevitable graviton bremsstrahlung associated with inflaton decay can be a unique probe of non-thermal leptogenesis. The emitted gravitons contribute to a high-frequency stochastic gravitational waves background with a characteristic fall-off below the peak frequency. Besides leading to a lower bound on the frequency  ($f\gtrsim 10^{11}$ Hz), the seesaw-perturbativity condition makes the mechanism sensitive to the lightest neutrino mass. For an inflaton mass close to the Planck scale, the gravitational waves contribute to sizeable dark radiation, which is within the projected sensitivity limits of  future experiments such as CMB-S4 and CMB-HD.

\end{abstract}

\maketitle

\section{Introduction}


Leptogenesis \cite{lep1} is an elegant mechanism to address the observed baryon asymmetry of the universe \cite{Planck}. The mechanism creates lepton asymmetry in the first step, which finally gets converted to baryon asymmetry via sphaleron-transition \cite{Kuzmin:1985mm}. There are several processes \cite{Alvarez-Gaume:1983ihn,Alexander:2004us,Caldwell:2017chz,Adshead:2017znw,Davoudiasl:2004gf,Lambiase:2006md,McDonald:2014yfg,McDonald:2015ooa} which generate lepton asymmetry in the Early Universe (EU).  The simplest one, perhaps, is the CP-violating and out-of-equilibrium decays of the right-handed (RH) neutrinos \cite{lep1, lep2,lep3,lep4,lep5,lep6,lep7} which are introduced in the Standard Model (SM) to generate light neutrino masses via Type-I seesaw. Broadly, there are two variants of leptogenesis in Type-I seesaw. I) Thermal leptogenesis, wherein thermal scatterings govern the fundamental dynamics of lepton asymmetry production \cite{lep1}. II) Non-thermal leptogenesis, which is not influenced by the thermal scatterings, and the RH neutrinos are produced from the decay of another field, such as inflaton \cite{nonlep1,nonlep2,nonlep3,nonlep4,nonlep5,nonlep6,Ghoshal:2022fud}. In this work, we discuss the latter.

Despite being elegant, leptogenesis is difficult to test, generally owing to the involvement of high energy scales beyond the reach of terrestrial experiments such as the Large Hadron Collider (LHC). There are proposals \cite{lep3, lowlep1,lowlep2,daSilva:2022mrx} to bring down the scale of leptogenesis, which, however, either yielded null results or await confrontation with future particle physics experiments. 

Indirect searches of high-scale leptogenesis are led mainly by neutrino observables at low energy, e.g., neutrino masses, mixing, CP-violating phases and the matrix element of neutrino-less double beta decay \cite{Altarelli:2010gt,rev1,rev2}. Alongside these, we might mention some contemporary and new tests of high-scale leptogenesis, which include, for instance, signatures from the meta-stability condition of Higgs vacuum in the early universe \cite{Ipek:2018sai, Croon:2019dfw}, and Cosmic Microwave Background Radiation (CMBR) measurements \cite{Ghoshal:2022fud}. 

In the new cosmic frontiers, the discovery of Gravitational Waves (GWs) from black hole mergers by LIGO and Virgo collaboration \cite{LIGOScientific:2016aoc,LIGOScientific:2016sjg} has encouraged us to put in efforts to detect GWs of primordial origins. Detection of primordial GWs would be of immense significance because GWs, unlike electromagnetic radiation, travel through the universe practically unimpeded with undistorted information about their origin. Therefore, they serve as the cleanest probe of physical processes, even at super-high energy scales.

Research on GWs of primordial origins at scales (wavelengths) smaller than CMB is now gaining considerable attention to test and constrain high-scale leptogenesis scenarios. This encases,  e.g., studies on the properties of cosmological phase transitions and their remnants. A catalog for these works, which might not be exhaustive, would include GWs from cosmic strings \cite{Dror:2019syi,Blasi:2020wpy,grav1,grav2,plep2}, domain walls \cite{Barman:2022yos}, plus nucleating and colliding vacuum bubbles \cite{Dasgupta:2022isg,Borah:2022cdx}. In addition, the stochastic backgrounds of GWs created by gravitons \cite{plep1} and cosmological perturbations 
 \cite{plep3} have also been investigated to search for the imprints of high-scale leptogenesis. Intriguingly, all these works reveal that the primordial GWs and their spectral features have prodigious potential to probe a wide variety of leptogenesis mechanisms, e.g., high-scale thermal leptogenesis \cite{Dror:2019syi}, gravitational leptogenesis \cite{grav1}, and leptogenesis induced by ultralight primordial black holes \cite{plep1}.

\begin{figure}
    \includegraphics[scale=.45]{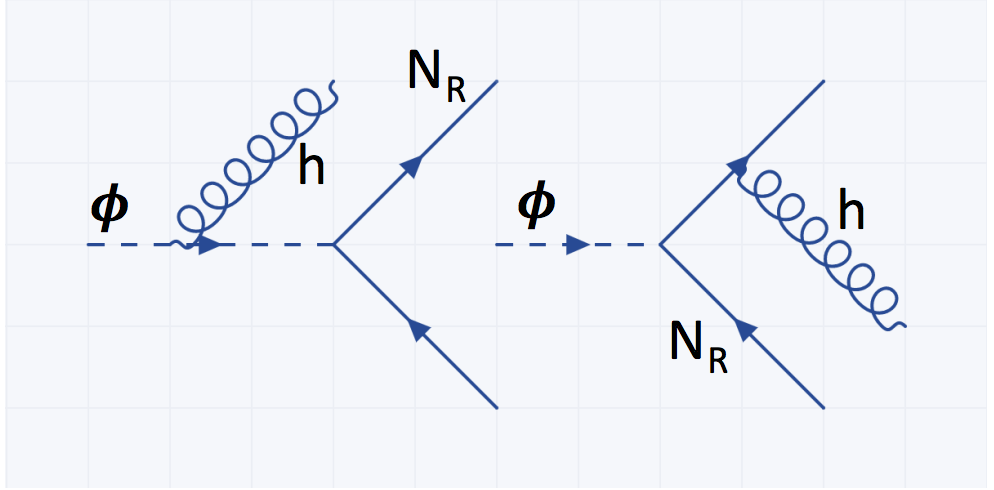}
    \caption{\it Diagrams representing the three-body decay of inflaton ($\phi$) to right-handed neutrinos (N$_R$) and graviton (h)  bremsstrahlung. A similar diagram  with a graviton attached to the incoming fermion line also contributes to the total decay-width. However, the four-point interaction vanishes \cite{Nakayama:2018ptw}.}
    \label{grav_fig}\label{fig1}
\end{figure}

We follow a similar line of research here; we study the possibility of testing non-thermal leptogenesis from inflaton decay with stochastic GWs constituting gravitons. We exploit the idea presented in Refs.\cite{Nakayama:2018ptw,Ema:2021fdz,Ema:2020ggo}, that after the end of inflation, once the coherent oscillation phase is over, the inflaton produces graviton bremsstrahlung by a three-body decay,  where the final states could be a couple of particles such as scalars and fermions (in our case, they are RH neutrinos) plus the graviton (see, e.g., Fig.\ref{fig1}). The gravitational waves constituting these gravitons contain the following features: I) Generally, the frequency of such GWs is very high. II) The GW spectrum is bounded from above and below. The high-frequency cut-off is typically set by the fraction of energy injected into the gravitons from the inflaton. In contrast, the low-frequency cut-off is determined by a threshold energy scale $\Lambda$, below which particle description of graviton is questionable. III) The GW spectrum exhibits a distinct fall-off below the peak frequency. IV) The peak frequency reaches its minimum value for the highest allowed  reheating temperature in the model. Finally, V) for inflaton mass close to the Planck scale, the energy density of the GWs increases so much that they contribute to testable dark radiation.

For non-thermal leptogenesis in Type-I seesaw,  first of all, the reheating temperature and the inflaton mass $m_\phi$ appear explicitly in the expression of lepton asymmetry. Therefore, they relate the leptogenesis parameter space with the properties of GWs. In addition, and most importantly, the effects of thermal scatterings mediated by the Yukawa interactions become negligible if $T_{RH}\lesssim M_R$, where $M_{R}$ is the RH neutrino mass scale. The seesaw-perturbativity condition ${\rm Tr}[f_N^\dagger f_N ]<4\pi$, with $f_N$ being the Yukawa coupling, implies that the RH neutrino mass scale has to be bounded from above to comply with the neutrino oscillation data on light neutrino masses -- this sets an upper bound on $T_{RH}$. We show how, in this way, and depending on the seesaw models, the lightest neutrino mass $m_1$ plays a crucial role in establishing a complementarity between the GW-physics and the physics of low-energy neutrinos to test and constrain non-thermal leptogenesis. \\

The rest is organised as follows: in Sec.\ref{s2}, we briefly describe the graviton emission from inflaton decay and the features of the expected GW background. In Sec.\ref{s3}, we discuss how non-thermal leptogenesis could be tested and constrained with such GWs. We summarise our results in Sec.\ref{s4}.

\section{Gravitational waves from Inflaton decay}\label{s2}
We closely follow Ref. \cite{Nakayama:2018ptw} to calculate the GW spectrum from the three-body decay of inflaton. The action leading to  such decays (see, e.g., Fig.\ref{fig1}) reads
\begin{widetext}
\begin{equation}
S= \int d^4x\sqrt{|g|}\left[\frac{M^2_P}{2}R+\frac{1}{2}g^{\mu\nu}\partial_\mu \phi \partial_{\nu} \phi - V(\phi) + \mathcal{L_{\rm int}}  \right], \label{eq:lag}
\end{equation}
\end{widetext}
where $M_{pl}\simeq 2.4\times 10^{18}$ GeV, $\phi$ is the inflaton with $V(\phi)$ being the potential and 
\begin{equation}
 \mathcal{L}_{\rm int} \supset  y_N \phi N_R N_R  
\end{equation} 
 accounts for the interaction of RH neutrinos with inflaton. To describe the decays with graviton emission,  it is sufficient to consider the effective interaction
\begin{equation}
	 \mathcal{L}^{\rm eff}_{\rm int} \supset \frac{\kappa}{2}h_{\mu\nu}T^{\mu\nu},
\end{equation}
where $\kappa =\sqrt{2}/M_{Pl}$, $h_{\mu\nu}$ is the graviton field defining the quantum fluctuation over the background, $g_{\mu\nu}=\eta_{\mu\nu}+\kappa h_{\mu\nu}$, and $T^{\mu\nu}$ is the energy-momentum tensor of the other fields. To compute the inflaton decay rate to the RH neutrinos but without the graviton, a flat spacetime background  $\eta_{\mu\nu}$ can be considered. In which case, the  two-body decay width of inflaton to RH neutrinos is given by 
\bea
\Gamma_0 (\phi \rightarrow N_R N_R ) &= \frac{ y_n ^2 m_\phi}{8\pi }\left(1-4y^2\right)^\frac{3}{2}, \label{2bd}
\eea
where $m_\phi$ is the mass of the inflaton, $y$ is a dimensionless parameter defined as $M_R/m_\phi$, and $M_R$ being the RH neutrino mass. On the other hand, the decay width involving gravitons is suppressed by a factor $m_\phi^2/M_{Pl}^2$, assuming the effective field theory holds for $m_\phi< M_{Pl}$. The contributions arising from such processes can be parametrized as \cite{Nakayama:2018ptw}
\bea
\Gamma_1\simeq \Gamma_0\left(\frac{1}{16\pi^2}\frac{m_\phi^2}{M_{pl}^2}\int_\Lambda^{m_\phi/2}\frac{dE}{E}\right), \label{f2bd}
\eea
where $E$ is the energy of the emitted gravitons and $\Lambda$ is a low energy threshold below which particle description is inaccurate. To compute $\Lambda$, we shall assume that the wavelength of the gravitons is shorter than the mean separation of the inflton particle, $n_\phi^{-1/3}$, where $n_\phi$ is the number density of inflaton. Assuming an instantaneous reheating, we calculate $\Lambda$ as
\bea
\Lambda\simeq\left[\rho_R (T_{RH})/ m_\phi\right]^{1/3},
\eea 
where $\rho(T_{RH})$ is the radiation energy density at the reheating temperature $T_{RH}$. The normalized energy density of the GWs that comes out from such decays is given by 
\bea
\Omega_{GW}\left(f\right)=\frac{1}{\rho_c}\frac{d\rho}{ d~{\rm ln}f},\label{GW1}
\eea
where $\rho_c=3M_{Pl}^2H_0^2$ is the critical energy density with $H_0=1.44\times 10^{-42}$ GeV being the Hubble parameter today, and $f$ is the present-day frequency of the GWs. \\

\begin{figure*}
    \includegraphics[scale=.315]{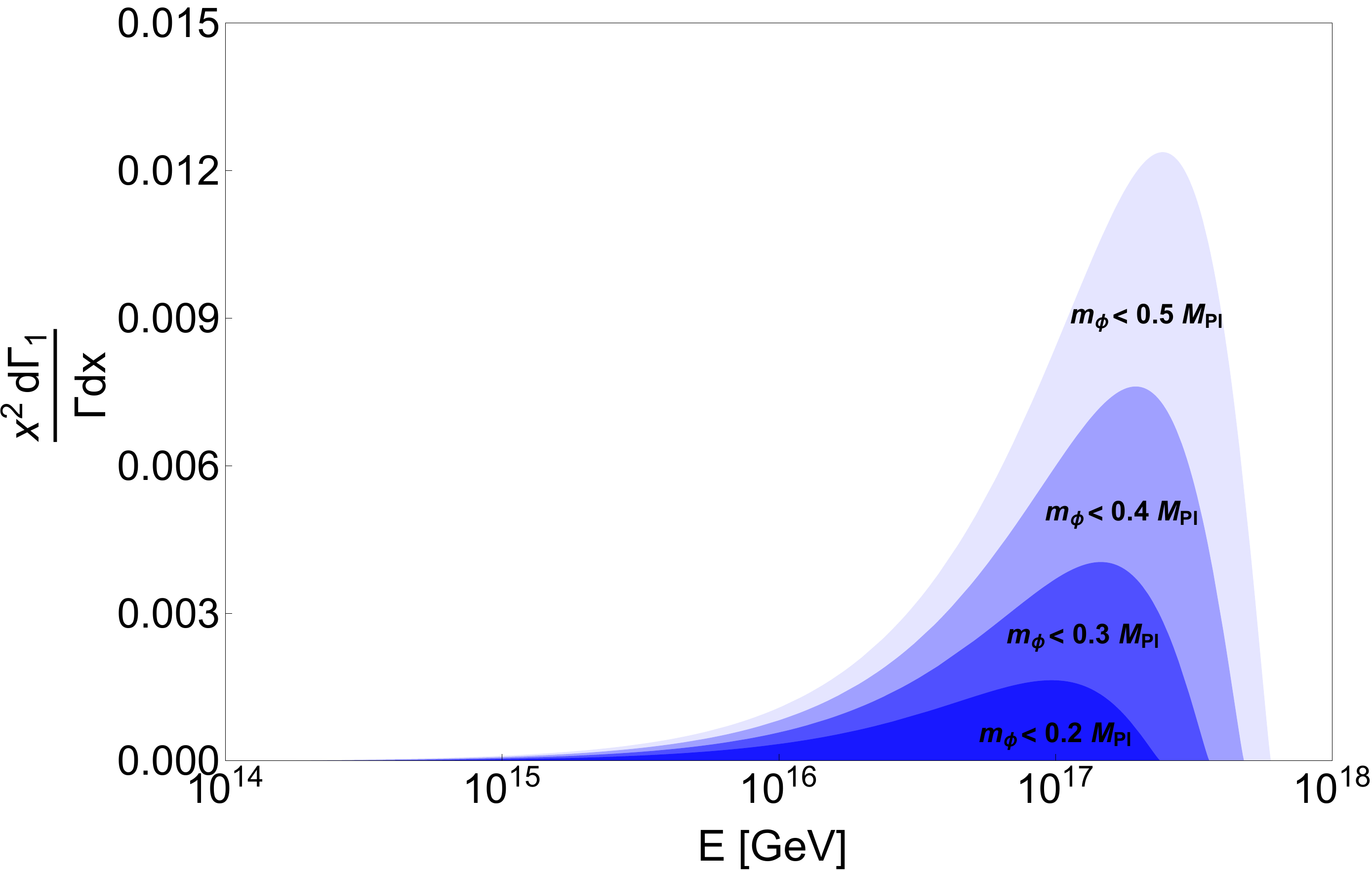}\includegraphics[scale=.315]{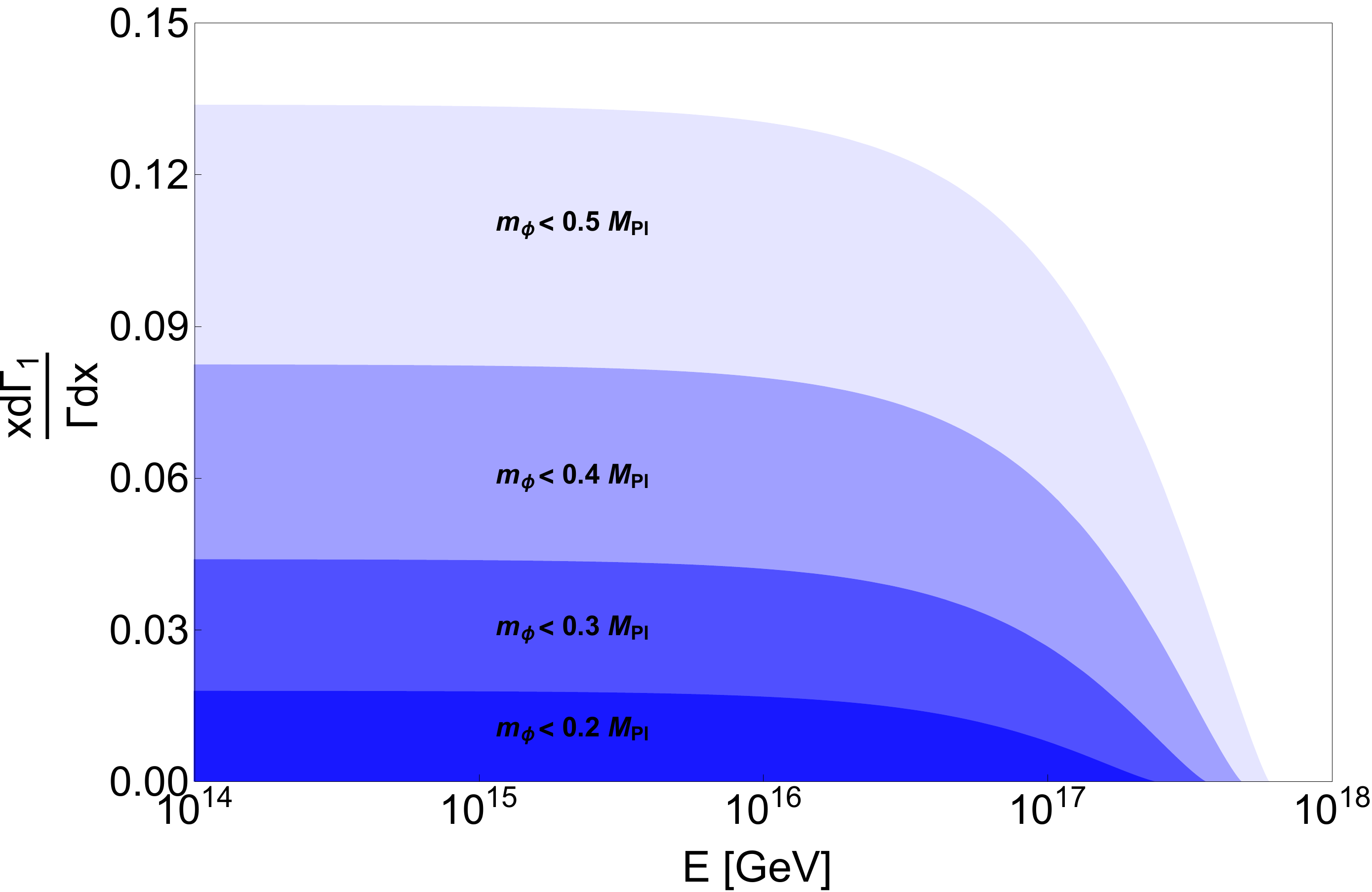}
    \caption{\it Form of the graviton energy spectrum with Left  $\frac{x^2 d\Gamma_1}{\Gamma dx}$ vs. $E$ (in GeV), Right: $\frac{x d\Gamma_1}{\Gamma dx}$ vs. $E$ (in GeV) for different values of $m_\phi$ and with the same benchmark values of the other parameters used in Fig.\ref{GWp}. When plotted in the Log-Log scale, the envelope of the upper region in the left panel gives the blue curve in the $\Omega_{GW}$ vs. $f$ graph in Fig.\ref{GWp}.}
    \label{dgvse}
\end{figure*}

Eq.\ref{GW1} can be re-expressed as (see derivation in appendix \ref{a1})
\bea
\Omega_{GW}\left(f\right)\simeq \frac{\rho_\gamma}{\rho_c}\left[\frac{2}{g_s\left(T_{RH}\right)}\right]^{1/3}\frac{\Gamma_1/\Gamma}{1-\bar{x}}\frac{x^2d\Gamma_1}{\Gamma_1 dx},\label{GW2}
\eea
where $\rho_\gamma/\rho_c\simeq 5.4\times 10^{-5}$, $\Gamma=\Gamma_0+\Gamma_1$, and $x=E/m_\phi=2\pi f \frac{T_{RH}}{m_\phi T_\gamma}$. The other two quantities $\bar{x}$ and $\frac{x^2 d\Gamma_1}{\Gamma_1 dx}$ (see, Fig.\ref{dgvse}, left) typically define the spectral shapes of the GWs. The total fraction of inflaton energy that is injected into the gravitons is given by 
\bea
\bar{x}=\frac{\Gamma_1}{\Gamma}\int_{\Lambda/m_\phi}^{1/2}\frac{x d\Gamma_1}{\Gamma_1 dx} dx.
\eea

For small values of $x$, the normalised graviton spectrum $\frac{x d\Gamma_1}{\Gamma_1 dx}$ (expression given in the appendix \ref{b1}) is constant (see, Fig.\ref{dgvse}, right), and as $x$ increases, the spectrum drops sharply. This feature has two important implications. First, most of the emitted gravitons are in the low energy range, and second, the spectrum goes as $\Omega_{GW}\sim f$ at low frequencies (cf. Eq.\ref{GW2}). The quantity $\bar{x}$ could be large enough ($\bar{x}\sim 10^{-2}$) when $m_\phi$ is close to the Planck scale, meaning that the highest frequency of the GWs could be as large as $0.01 M_{Pl}T_\gamma/T_{RH}$ which for $T_{RH}\sim 10^{15}$ GeV, gives $f_{\rm high}\sim 10^{14}$ Hz.

\section{Non-thermal leptogenesis and gravitational waves}\label{s3}

The minimal Lagrangian that facilitates light neutrino masses and leptogenesis is given by 
\begin{equation}
-\mathcal{L}^{\rm seesaw}= f_{N\alpha i}\bar{\ell}_{L\alpha}\tilde{\eta} N_{Ri}
+\frac{1}{2}\bar{N}_{Ri}^CM_{ij} \delta _{ij}N_{Rj} 
+ {\rm h.c.}\,, \label{seesawlag}
\end{equation}
where $\ell_{L\alpha}=\begin{pmatrix}\nu_{L\alpha} & e_{L\alpha}\end{pmatrix}^T$ is the SM lepton doublet of flavor $\alpha$, $\tilde{\eta}=i\sigma^2 \eta^*$ with $\eta= \begin{pmatrix}\eta^+&\eta^0
\end{pmatrix}^T $ being the Higgs doublet with $\braket{\eta_0}\equiv v = 174$ GeV and $M={\rm diag}\hspace{.5mm} (M_1,M_2,M_3)$, $M_{1,2,3}>0$. After Electroweak symmetry breaking, the light neutrino mass matrix 
\bea
M_\nu = -m_DM^{-1}m_D^T=U D_m U^T \label{seesawfor}
\eea
is obtained via  standard seesaw mechanism \cite{Minkowski:1977sc,Gell-Mann:1979vob,Yanagida:1980xy,Mohapatra:1979ia}. In Eq.\ref{seesawfor}, the matrices $m_D=f_N v$ is the Dirac neutrino mass matrix, $D_m=-~{\rm diag}~(m_1,m_2,m_3)$ with $m_{1,2,3}>0$ being the physical light neutrino masses and $U$ is the  $U_{PMNS}$ matrix that mixes the flavour and mass eigenstates of the light neutrinos.
The $B-L$ asymmetry created by the CP-violating and out-of-equilibrium decays of RH neutrinos is given by \cite{lep4}
\bea
N_{B-L}=\sum_{i}^3\varepsilon_{i}\kappa_i\,, \label{bau0}
\eea
where  $\varepsilon_i$  and $\kappa_i$ are the unflavoured CP asymmetry parameter, and the efficiency of the asymmetry production corresponding to the $i$th RH neutrino. The efficiency factor $\kappa_i$, which takes into account the combined effects of asymmetry production and washout, is given by \cite{lep4}

\begin{equation}
\kappa_i \left(z=\frac{M_1}{T}\right)=-\int_{z_{\rm T_{RH}}}^z \frac{dN_{N_i}}{dz^\prime}e^{-\sum_{i}\int_{z^\prime}^z W_{i}^{\rm ID}(z^{\prime\prime})dz^{\prime\prime}}dz^\prime\,.
\label{effi1}
\end{equation}

The number densities in Eq.\ref{bau0} and Eq.\ref{effi1}, have been normalised by the ultra-relativistic number density of $N_i~(=g_{N_i}T^3/\pi^2)$, where $g_{N_i}=2$ . In the case of thermal leptogenesis, the temperature in the EU is high enough ($T>M_i$) to populate RH neutrino number densities by Yukawa scatterings and to facilitate strong washout effects that erase a significant lepton asymmetry. The inverse decay term $W_{i}^{\rm ID}=\Gamma_{i}^{\rm ID}/Hz$, which takes into account the washout effect, is given by
\bea
W_i^{\rm ID}=\frac{1}{4}K_i\sqrt{r_{1i}}\mathcal{K}_1(z_i)z_i^3\,,\label{inv_decay}
\eea
where $H$ is the Hubble parameter, $r_{ij}=M_j^2/M_i^2$, $\mathcal{K}_1$ is a modified Bessel's function, and $K_i$ is the decay-parameter corresponding to $i$th RH neutrino, defined as
\beq
K_i= \frac{(m_D^\dagger  m_D)_{ii}}{M_i m^*},
\eeq
with  $m^*\simeq10^{-3}$ being the equilibrium neutrino mass \cite{lep3,lep4}.
A leptogenesis scenario sensitive to inflationary physics requires less efficient thermal scatterings so that the thermal production of RH neutrinos, which generally erase the initial conditions, is negligible.  This happens if $T_{\rm RH}\lesssim M_{i}$, and we can also neglect the  washout effects ($W_{i}^{\rm ID}$)\footnote{The condition $z\gtrsim z_{RH}$ corresponds to $\Gamma_0/H\gtrsim 1$. The inverse decays go out-of-equilibrium, i.e., $\Gamma_{\rm ID}/H\lesssim 1$, at $z\gtrsim z_{out}$. Therefore, $z_{RH}\gtrsim z_{out}$ implies $\Gamma_0/H\gtrsim \Gamma_{\rm ID}/H$. Generally, $z_{out}$ is taken to be 1. Consequently, the condition for non-thermal leptogenesis becomes; $z_{RH}=M/T_{RH}\gtrsim 1$, i.e., $M\gtrsim T_{RH}$. Nonetheless, $z_{out}$ is modulated by Yukawa couplings in a way \cite{lep4} that $z_{out}=\Phi(K)>2$, -- $\Phi(K)$ has been introduced in the main text. Therefore, a more accurate condition for non-thermal leptogenesis reads $M\gtrsim \Phi(K)T_{RH}$, which we maintain throughout the article. }. The simplest source of non-thermal production of RH neutrinos is a tree-level decay of inflaton \cite{nonlep1,nonlep2}. Assuming inflaton decays to all the RH neutrinos with the same branching fraction $B_N$, and the RH neutrinos instantaneously decay to produce lepton and anti-lepton states, an expression for the total lepton asymmetry can be derived as \cite{Samanta:2020gdw}
\begin{equation}
N_{B-L}=\sum_i N_{N_i}\biggr\rvert_{\rm RH} \varepsilon_i\equiv\frac{\pi^4 B_N g_{*}}{30 g_N}\sum_i \frac{\varepsilon_i T_{\rm RH}}{ m_\phi} \label{bau}.
\end{equation}
Successful baryogenesis via leptogenesis then requires \cite{Planck}
\bea
\eta_B=a_{\rm sph}\frac{N_{B-L}}{N_\gamma^{\rm rec}}\simeq 0.96\times 10^{-2}N_{B-L}\noindent
\\ \simeq \eta_B^{CMB}=\left(6.3\pm 0.3\right)\times10^{-10},
\eea 
where $N_\gamma^{\rm rec}$ is the normalized photon density at the recombination and the sphaleron conversion coefficient $a_{\rm sph}\sim 1/3$. A more accurate lower bound on $M_i$ can be derived by generalising the condition $M_i>T_{\rm RH}$ to $M_i>\Phi(K_i)T_{\rm RH}$. The function $\Phi(K_i)$ determines the exact value of $z_f=M_i/T$ at which washout processes go out of equilibrium. The function $\Phi(K_i)$ can be calculated analytically \cite{lep4}, and is given by 
\bea
\Phi\left( K_i\right)=2+4  K_i^{0.13} e^{\frac{-2.5}{ K_i}}.\label{phik}
\eea

It is convenient to parametrize \cite{Casas:2001sr} the Dirac mass matrix as 
\bea
m_D=U\sqrt{D_m}\Omega\sqrt{M_R},\label{orth}
\eea
where $\Omega$ is a $3\times 3$ complex orthogonal matrix given  by 
\begin{widetext}
\begin{equation}
\Omega = 
\begin{pmatrix}
1 &  0 & 0 \\
0 & \cos z_{23} & \sin z_{23} \\
0 & - \sin z_{23}  & \cos z_{23}
\end{pmatrix}  
\begin{pmatrix}
\cos z_{13} & 0  & \sin z_{13} \\
0 & 1 & 0 \\
- \sin z_{13} & 0 & \cos z_{13}
\end{pmatrix}  
\begin{pmatrix}
\cos z_{12} & \sin z_{12} & 0 \\
-\sin z_{12} & \cos z_{12} & 0 \\
0 & 0 & 1
\end{pmatrix}
\end{equation}
\end{widetext}
with $z_{ij}=x_{ij}+i y_{ij}$. The orthogonal parametrization has two important aspects. First, it shows that non-thermal leptogenesis is not sensitive to neutrino oscillation experiments (mixing angles and low-energy CP phases) because $\varepsilon_i\propto {\rm Im}\left[(m_D^\dagger m_D)_{ij}\right]$ which is independent of $U$. Therefore, the CP asymmetry is stemmed only from the complex phases within $\Omega$. Second, it helps us to understand the relation among the  states ($\ell$) produced by the heavy neutrinos  and light neutrino states ($\tilde{\ell}$) as
\bea
\ket{\ell_j} =B_{ji}\ket{\tilde{\ell}_i},
\eea
where the `bridging matrix' $B_{ij}$ \cite{DiBari:2018fvo},  is given by
\bea
B_{ji}=\frac{\sqrt{m_i}\Omega_{ji}}{\sqrt{m_k|\Omega_{kj}|^2}}.\label{bri}
\eea

\begin{figure}
\includegraphics[scale=.4]{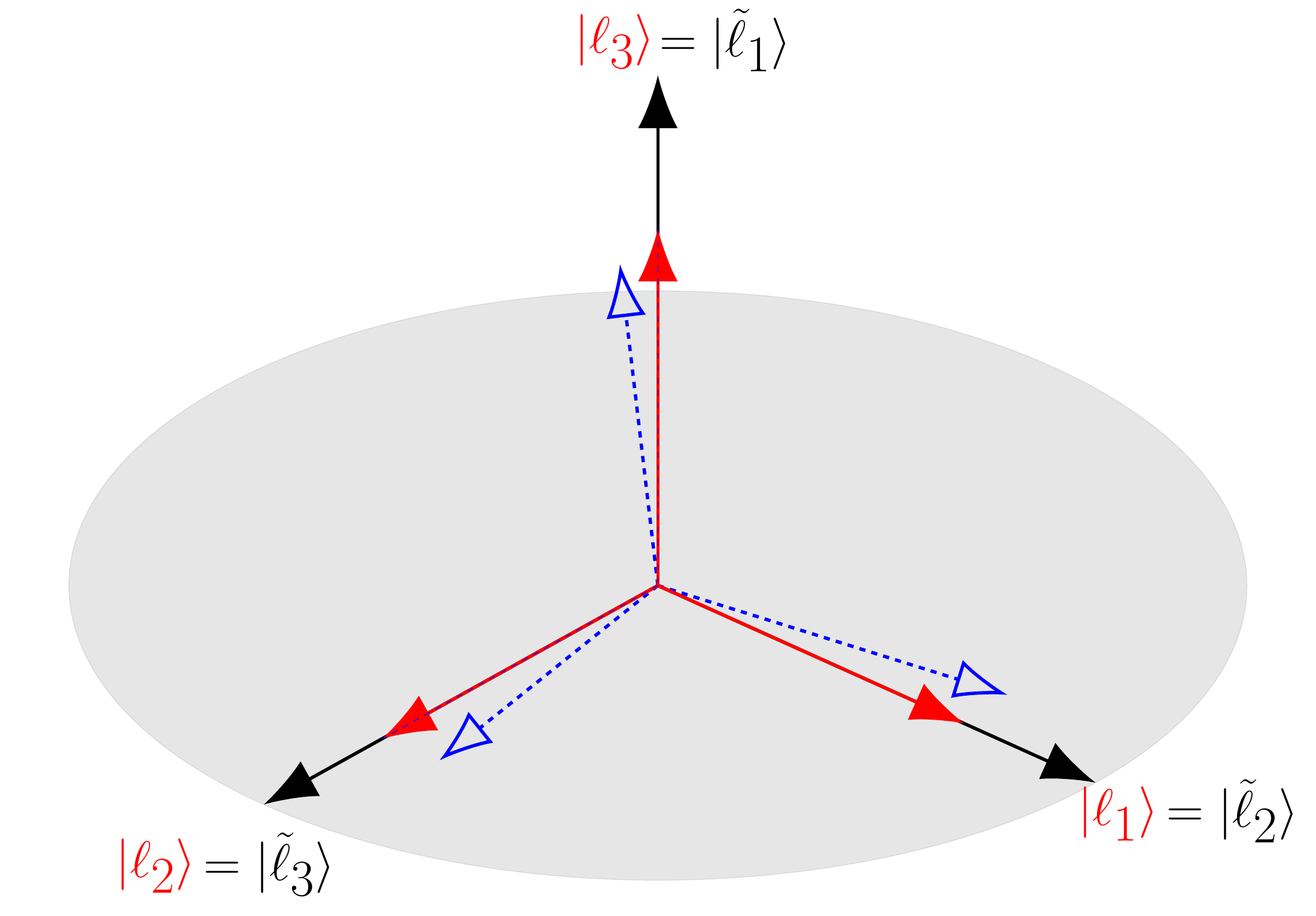}
\caption{\it Illustration of seesaw models: The black arrows represent the light neutrino state vectors, whereas the  state vectors in red are the heavy states produced by the RH neutrinos when the orthogonal matrix is a permutation matrix. The configuration of the heavy states shown by the blue arrows corresponds to an arbitrarily complex orthogonal matrix.} \label{bases}
\end{figure}
If the orthogonal matrix is a permutation matrix (could be identity, in the case no permutation) \cite{Chen:2009um},  the heavy and the light states coincide (Fig.\ref{bases} black and red vectors). Note also that this configuration is unable to generate CP asymmetry because $\Omega$ is real. On the other hand, for an arbitrarily complex $\Omega$, the light and heavy states do not coincide (Fig.\ref{bases} black and dashed-blue vectors). Due to its complex nature, $\Omega$ can have large entries, and the seesaw model is said to be fine-tuned \cite{DiBari:2018fvo}. The fine-tuning parameter, $\gamma_i=\sum_j |\Omega^2_{ij}|\geq 1$  accounts for the fractional contribution of the heavy states ($M_j$) to a  light state ($m_i$). For larger  entries in the imaginary parts of $\Omega$, the heavy states (e.g., blue vectors in Fig.\ref{bases}) disperse\footnote{Because $\Omega$ belongs to $SO(3,\mathbb{C})$, it is isomorphic to the Lorentz group. Therefore, $\Omega$ can be factorized as  $\Omega = \Omega^{\rm rotation}\Omega^{\rm Boost}$. For large entries in the $\Omega^{\rm Boost}$ matrix, the seesaw model gets more fine-tuned or `boosted,' $\gamma_i=\sum_j |\Omega^2_{ij}|\gg 1$.} more from the light states. In this article, we shall present results for small (minimally fine-tuned), intermediate (moderately fine-tuned), and large (extremely fine-tuned) values of $\gamma_i$. \\

Another important constraint that must be considered is the seesaw-perturbativity condition ${\rm Tr}\left(f_N^\dagger f_N\right)\leq 4\pi$. For a single-scale seesaw $(M_{i=1,2,3}\equiv M)$, which we adopt in this work, the condition reads

\bea
M\lesssim\frac{4\pi v^2}{m^*\sum_i K_i}\lesssim \frac{4\pi v^2}{\sum_i \sum_km_k |\Omega_{ki}|^2}.
\eea

For quasi degenerate RH neutrinos the function $\Phi(K_i)$ gets generalised to  $\tilde{\Phi}\equiv\Phi(\sum _i K_i)$ \cite{DiBari:2019zcc}, therefore, the upper bound on the reheating temperature becomes
\bea
T_{RH}\lesssim  \frac{4\pi v^2}{\Phi(\sum _i K_i) \sum_i \sum_km_k |\Omega_{ki}|^2 }.
\eea
\\

\begin{figure*}
\includegraphics[scale=.32]{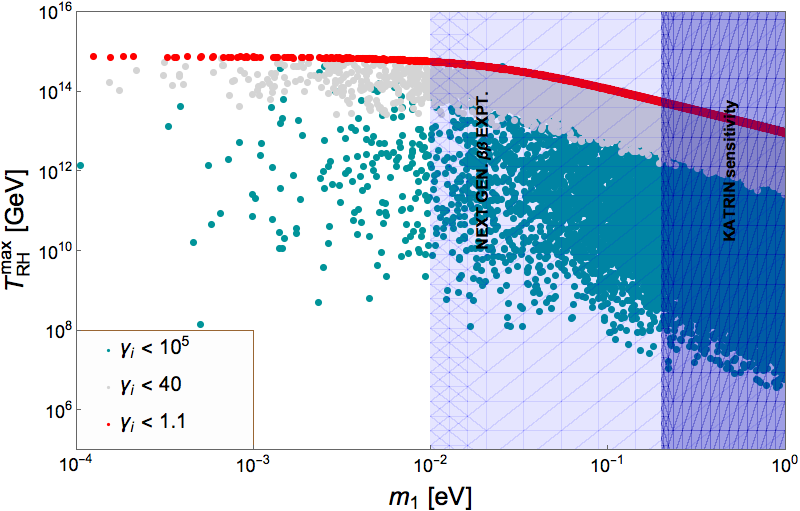}\includegraphics[scale=.32]{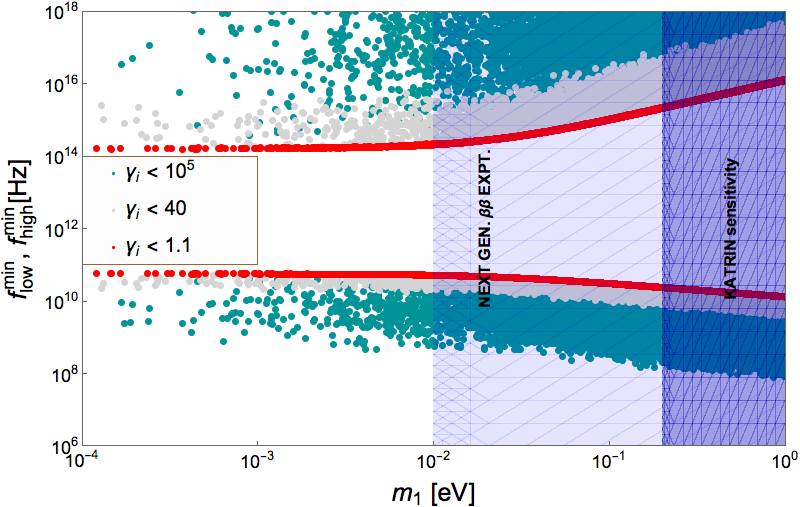}
\caption{\it Left: scatter plot of $T_{RH}$ vs. $m_1$:  $\gamma_i\leq 1.1$ (red), $\gamma_i\leq 40$ (gray), $\gamma_i\leq 10^5$ (green). We randomly varied the complex angles in the orthogonal matrix to achieve the desired values of $\gamma_i$. The light neutrino masses are generated taking into account the solar and atmospheric mass squared differences from neutrino 3$\sigma$ oscillation data \cite{rev2} and varying $m_1$ randomly within the interval $10^{-4}$ eV-1 eV. Right: scatter plot of $f_{\rm peak, low}^{\rm min}$, $f_{\rm peak, high}^{\rm min}$ vs. $m_1$ for $m_\phi=0.5 M_{pl}$ giving us the correlation between GW and low energy neutrino physics.} \label{trhm1}
\end{figure*}

In Fig.\ref{trhm1} (left), we have shown the variation of $T_{RH}$ with the lightest neutrino mass $m_1$ for different values of $\gamma_i$. A couple of crucial points can be extracted from this plot. First, as $m_1$ increases, the upper bound ($T_{RH}^{\rm max}$) on the $T_{RH}$ decreases. This is true for all the choices of $\gamma_i$. Second, the parameter space enlarges for larger values of $\gamma_i$ to include lower values of $T_{RH}^{\rm max}$. Therefore, as the seesaw models exhibit more boosted configurations (large $\gamma_i$), non-thermal leptogenesis happens for lower values of $T_{RH}$.  \\

\begin{figure*}
\includegraphics[scale=.33]{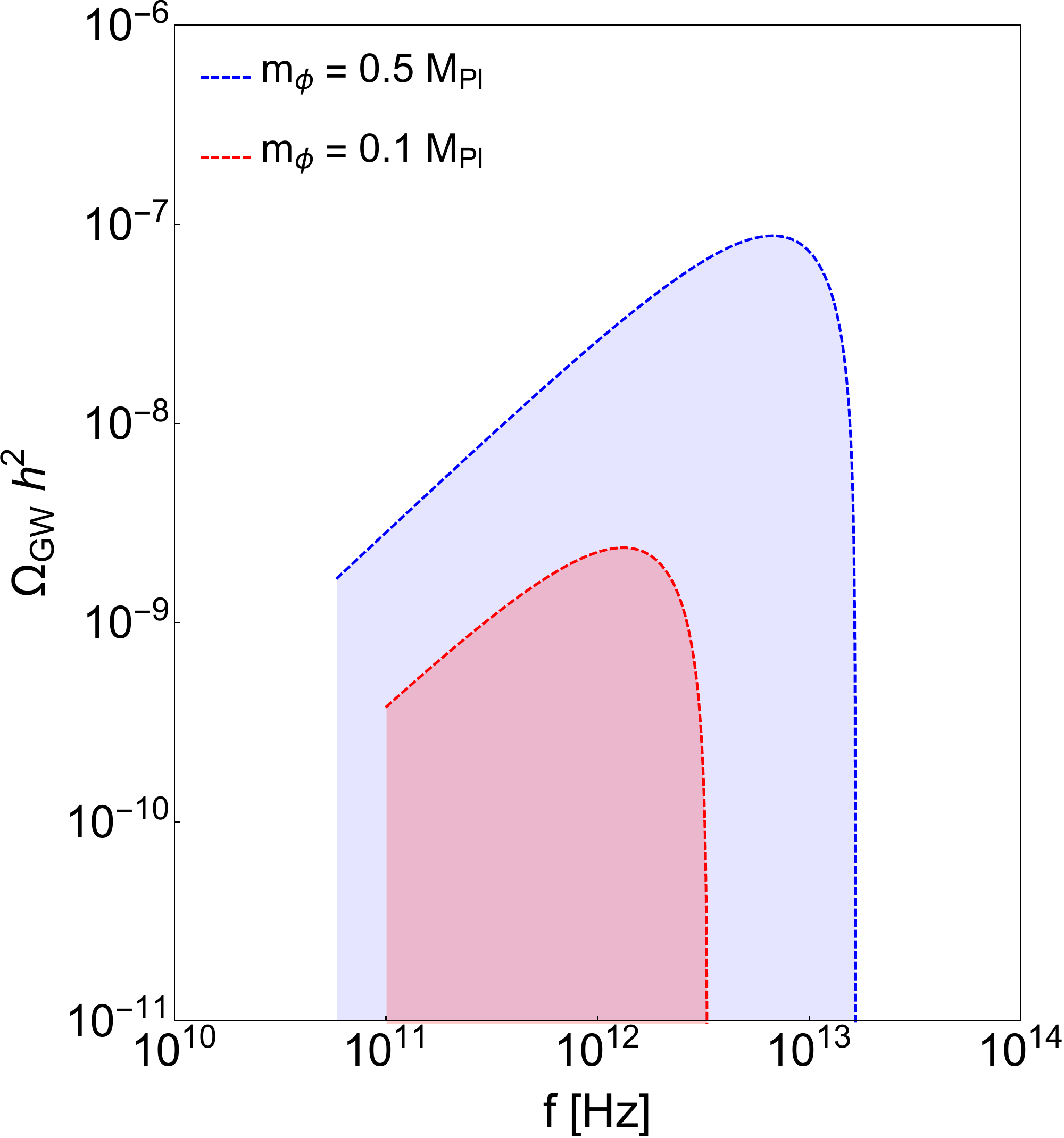}\includegraphics[scale=.52]{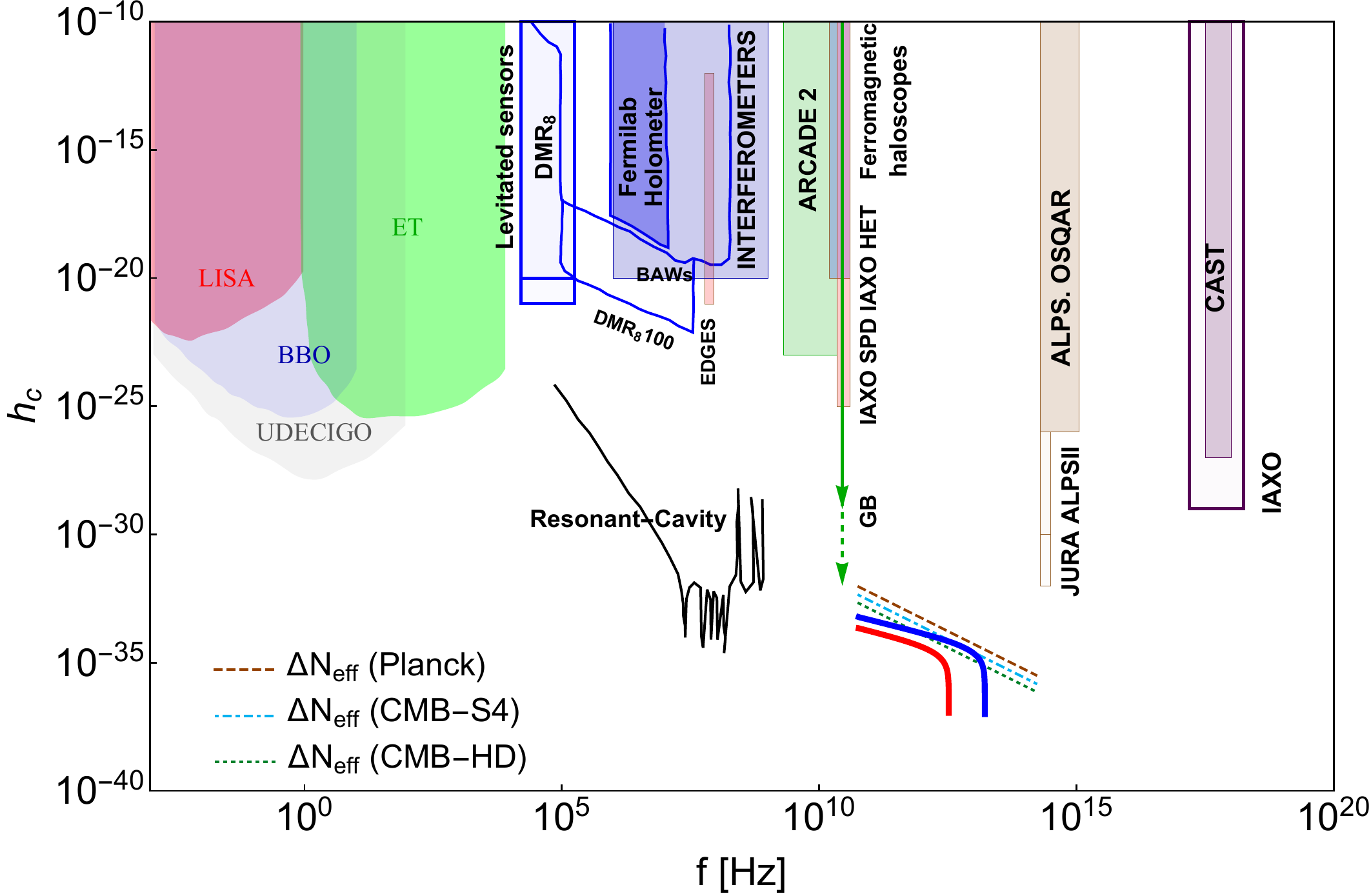}
\caption{\it Left: $\Omega$ vs. $f$ plot for $m_\phi \lesssim 0.5M_{Pl}$ and $m_\phi\lesssim 0.1 M_{Pl}$. For the $m_\phi \simeq 0.5M_{Pl}$ curve, relevant parameters to generate correct baryon asymmetry are mentioned in the text. To do the same for $m_\phi \simeq 0.1M_{Pl}$, one needs to adjust the parameters so that a factor $\sim 5$ appears in the denominator of Eq.\ref{bau}.  Right: $h_c$ vs. $f$ plot for the same values of $m_\phi\simeq 0.5,0.1 M_{Pl}$, subjected to the projected sensitivity limits of various high-frequency GW detectors  \cite{d1,d2,d3,d4,d5,d5a,d6,d7,d8,d9,d10,d11,d12,d13,d14,d15}. The dashed line is the current upper bound on $\Delta N_{\rm eff}$. The dot-dashed and the dotted lines are the future projections on the same. {\rm CMB-HD} has the potential to probe inflaton mass $m_\phi \sim 0.25 M_{Pl}$. Ref.\cite{cavity} suggests that the electromagnetic resonant cavities  could potentially reach below the dark radiation projections with strain $h_c\lesssim 10^{-30}$ within the frequency range $(10^6-10^8)~\rm Hz$. }\label{GWp}
\end{figure*}

We now proceed to the computations of the $f_{\rm low}$ and $f_{\rm high}$ and the discussion of possible complementary searches  in low-energy neutrino experiments. Because the energy of the produced gravitons will be red-shifted to the present time, the frequencies $f_{\rm low}$ and $f_{\rm high}$ are simply given by\footnote{The peak frequency can be well-approximated as $f_{\rm peak}\simeq \frac{m_\phi}{10\pi}\frac{T_\gamma}{T_{RH}}$, see appendix \ref{b1}.}

\begin{equation}
f_{\rm low}=\frac{\Lambda}{2\pi}\frac{T_\gamma}{T_{RH}},~~f_{\rm high}=\frac{m_\phi}{4\pi}\frac{T_\gamma}{T_{RH}}. \label{frq}
\end{equation}
Therefore, when the reheating temperature is maximum, we have
\begin{widetext}
\begin{equation}
f_{\rm low}^{\rm min}=\frac{\Lambda T_\gamma\left(\tilde{\Phi} \sum_i \sum_km_k |\Omega_{ki}|^2 \right)}{8\pi ^2 v^2},~~f_{\rm high}^{\rm min}=\frac{m_\phi T_\gamma\left(\tilde{\Phi}\sum_i \sum_km_k |\Omega_{ki}|^2 \right)}{16\pi^2 v^2}.
\end{equation}
\end{widetext}
In Fig.\ref{trhm1} (left), we have shown the variation of  $f_{\rm low}^{\rm min}$ and $ f_{\rm high}^{\rm min}$ with the lightest neutrino mass $m_1$ for different values of $\gamma_i$. In this context, let us point out the current experimental fact file of the lightest neutrino mass $m_1$. The sum of the light neutrino masses is bounded from above;  $\sum_im_i\lesssim$ 0.17 eV \cite{Planck}-- this corresponds to $m_1\lesssim$ 50 meV. A more stringent upper bound $m_1\lesssim$ 31 meV may be obtained from the latest PLANCK data \cite{sunny}. The dark-blue vertical region in Fig.\ref{trhm1} is the projected sensitivity of the KATRIN experiment, which is starting to measure neutrino masses with sensitivity to 0.2 eV \cite{Giuliani:2019uno}. Assuming a normal mass ordering $m_3>m_2>m_1$, a future discovery by the ongoing and the planned  neutrino-less $\beta\beta$ decay experiments \cite{Giuliani:2019uno} would correspond to $m_1\gtrsim 10^{-2}$ eV (light-blue vertical region in Fig.\ref{trhm1}).  Consider now seesaw models with an orthogonal matrix with small entries, e.g., $\gamma_i<1.1$ and the corresponding  $f-m_1$ curves, shown with red lines (Fig.\ref{trhm1}, right). In which case, a future discovery of GWs with $10^{11}{\rm Hz}<f_{\rm peak}\lesssim 10^{14}$ Hz, if that is really possible, would be in tension with $m_1\gtrsim10^{-2}$ eV. In other words, a future discovery of the lightest neutrino mass $m_1\gtrsim 10^{-2}$ eV would imply a non-thermal leptogenesis scenario will be associated with a broader spectrum of the GWs with the peak shifted to the higher frequencies than what is expected for the $m_1\lesssim 10^{-2}$ case.  In Fig.\ref{GWp}, we have shown $f$ vs. $\Omega_{GW}$ (left)  and $f$ vs. $h_c$ (right) curves which  correspond to successful leptogenesis (blue curves) for $x_{ij}\sim \pi/4$, $y_{ij}\sim 10^{-7}$, $m_1\sim 10^{-3}$, $m_\phi=0.5 M_{Pl}$, $B_N\simeq 1$, and for mass degeneracy among the heavy neutrinos $\delta\sim 10^{-1}$. We plotted another curve (in red) in both figures to show the inflaton mass dependence. To obtain these plots, we first calculate  $\Omega_{GW}$ as

\begin{equation}
\Omega_{GW}=\Omega_{GW}^0\left[\frac{2}{g_s\left(T_{RH}\right)}\right]^{1/3}\frac{\omega}{1+\omega}\frac{x^2}{1-\bar{x}}\frac{d\Gamma_1}{\Gamma_1 dx},\label{GWexp}
\end{equation}

where $\Omega_{GW}^0\simeq 5.4\times 10^{-5}$, the quantities $x$, $\bar{x}$, $\frac{d\Gamma_1}{dx}$ have the usual definitions (see sec.\ref{s2}), and we use
\begin{widetext}
\begin{equation}
\omega=\frac{1}{16\pi^2}\frac{m_\phi^2}{M_{pl}^2}\int_\Lambda^{m_\phi/2}\frac{dE}{E},~~\Lambda=\left[\frac{\pi^2g_s(T_{TH})T_{RH}^4}{30 m_\phi}\right]^{1/3}\hspace{-.6cm},~~\Gamma_1=\frac{\Gamma \omega}{1+\omega}
\end{equation}
\end{widetext}
with $\Gamma \simeq 11 T_{RH}^2/M_{Pl}-$assuming an instantaneous reheating. We finally calculate the dimensionless strain $h_c(f)$ as
\bea
h_c(f)=\sqrt{\frac{3H_0^2}{2\pi^2}\Omega_{GW}}f^{-1}.
\eea
The key takeaway points from these figures are,  the plotted curves represent the strongest GWs at the lowest allowed frequencies. This is because, for the used set of parameters, one   typically has the maximal reheat temperature, and therefore the spectrum peaks at $f_{\rm peak }=f_{\rm peak }^{\rm min}$. Once $T_{RH}$ decreases, the spectrum broadens with peaks  at higher frequencies and with much-reduced amplitudes at lower frequencies. Even though the GWs expected from this model can not be detectable directly (see the present constraints and projected limits \cite{d1,d2,d3,d4,d5,d5a,d6,d7,d8,d9,d10,d11,d12,d13,d14,d15} on high-frequency GWs in Fig.\ref{GWp}),  for inflaton mass close to Planck scale \footnote{It is not trivial to accommodate inflaton mass close to the Planck scale in the simple single-field scenarios. Nonetheless, in the models that  seek to generate all physical scales, including the Planck scale dynamically, it can be done. Based on scale invariance, these models generally deal with multi-field configurations (sometimes with sub-Planckian scalar mass eigenstates, which could be inflaton), which drive inflationary dynamics. Therefore a possible UV embedding of our scenario would go along the line of, e.g., \cite{ps1,ps2,ps3,ps4,ps5}.}, as we have chosen here, can produce significant dark radiation (DR) which is detectable in future experiments such as CMB-S4 \cite{CMB-S4:2016ple} and CMB-HD \cite{CMB-HD:2022bsz}. The dark radiation constraint on GWs reads
\bea
\int f^{-1} \Omega_{GW} h^2\left(f\right)df\lesssim 5.6\times 10^{-6}  \Delta {\rm N}_{\rm eff}.\label{neff}
\eea
For smaller values of $x=E/M$, i.e., in the low frequency range, the quantity $\frac{xd\Gamma_1}{\Gamma_1 dx}$ becomes constant (see Fig.\ref{dgvse}) and therefore  the GW spectrum below the $f_{\rm peak}$ goes as $\Omega_{GW}\sim f$ (cf. Eq.\ref{GWexp}). Consequently, we can parametrise the $\Omega_{GW}$ as $\Omega_{GW}\left(f\right)\sim \Omega_{GW}\left(f_{\rm low}\right)\left(\frac{f}{f_{\rm low}}\right)$ for $f\lesssim f_{\rm peak}$. Using Eq.\ref{neff}, we estimate that for $m_\phi =0.5 M_{Pl}$, one can have $\Delta {\rm N}_{\rm eff}$ as large as $0.04$ which is accessible in the future CMB-S4/HD measurement.  In this way, graviton bremsstrahlung of inflaton could be an interesting way to test and constrain non-thermal leptogenesis, {\it provided that the inflaton mass is close to} $M_{Pl}$.\\

One may naively get effective $m_\phi$ to be $\sim 10^{13}$ GeV for $m^2 \phi ^2$ inflation, which is ruled out as a model by $n_s/r$ measurements. However, one may always introduce a non-minimal coupling $\xi \phi^2 R$ like the Higgs inflation scenario to rescue such scenarios \cite{Tenkanen:2017jih}. As usual, the $\xi \phi^2 R$ term becomes irrelevant during and post the re-heating era, because the Ricci scalar {\rm R} is very close to 0.\\





We end this section with the following remarks:\\

 $\bullet$ Here we consider the total decay width of the inflaton $\Gamma=\Gamma_0+\Gamma_1$ (see below Eq.\ref{GW2} and the derivation of GWs in Appendix \ref{a1}), where $\Gamma_0$ and $\Gamma_1$ are defined in Eq.\ref{2bd} and Eq.\ref{f2bd}, respectively. Therefore, the scenario assumes that the decay products of RH neutrinos reheat the universe. Generally, in non-thermal leptogenesis models, this is a simplified assumption--so-called the RH neutrino reheating scenario, see, e.g., Refs. \cite{nonlep2,nonlep4,nonlep5,nonlep6}. One can redo a similar exercise by adding more channels, e.g., allowing direct inflaton coupling to SM particles. In that case,  branching ratios, i.e.,  the direct couplings of SM particles to inflaton, would play a crucial role. Therefore, it is not apparent that RH neutrino-bremsstrahlung would dominate in such a case. \\

$\bullet$ Models with specific flavor structures (with a definite $\gamma_i$) would be more predictive (cf. Fig.\ref{trhm1}). Though for simplicity, we discuss the non-thermal leptogenesis with quasi-degenerate RH neutrinos, we do not expect any qualitative difference for the hierarchical scenarios. For the latter, one has to impose the condition $T_{RH}\lesssim M_{\rm lightest}$. Therefore, if the mass spectrum is strongly hierarchical $M_3\gg M_2\gg M_{1, \rm lightest}$, the peaks of the GWs would shift to the higher frequency values.\\

$\bullet$ We mostly focus on high reheating temperatures here. It is evident from Eq.\ref{frq}, that for low reheating temperatures, the GW spectrum shifts towards high-frequency values. In which case, the produced graviton would be inevitably converted to photons in the presence of a cosmological magnetic field via inverse-Gertsenshtein effect \cite{igef}. Therefore, a non-thermal leptogenesis mechanism with a low-reheating temperature might produce an observable cosmic X-ray background \cite{Dolgov:2012be}. A detailed study in direction will be presented in a future publication.\\ 

$\bullet$ Such high-frequency gravitational waves may arise due to the graviton production from inflaton oscillation \cite{infosc1,infosc2}. Compared to the three-body decays, the relative magnitude of such a process can be derived assuming instantaneous reheating and $m_\phi\gg T_{RH}$ as
\begin{equation}
\mathcal{A}=\frac{\Gamma_{osc}}{\Gamma_1}\simeq \frac{T_{RH}^2}{7\times 10^{-4} m_\phi M_{Pl}}\left[{\rm ln}\left(\frac{m_\phi}{T_{RH}}\right)\right]^{-1},
\end{equation}
which for our preferred values of parameters is much less than unity ($\mathcal{A}\ll 1$).\\

$\bullet$ Hawking radiation and primordial density perturbations from ultralight PBH sources high-frequency gravitational waves \cite{hw1,hw2,hw3,hw4,Gehrman:2022imk}. Moreover, these PBHs can also produce RH neutrinos, which decay to create baryon asymmetry via leptogenesis \cite{plep1,plep2,plep3,Fujita:2014hha}. Therefore, this mechanism is  analogous to the present scenario (a setup to study non-thermal leptogenesis and gravitational waves). Nonetheless, the spectral features of the GWs produced by the PBHs are different \cite{Inomata:2020lmk} from what we discuss in this work. Additionally, depending on the PBH's production and distribution, one may have GWs at low frequencies \cite{plep3}, which distinguish it  from the inflaton decay scenario. \\

$\bullet$ Plenty of efforts and proposals have been put forward to detect high-frequency GWs \cite{d7,d8,d9,d10,d11,d12,d13,d14,d15}. This is because, apart from the graviton bremsstrahlung discussed in this paper and the sources discussed above, several other well-motivated sources produce high-frequency GWs. This includes, e.g., inflation \cite{hinf1,hinf2,hinf3}, pre-heating \cite{pre1,pre2,pre3} cosmic strings from a very high energy phase transition \cite{cs1,cs2,cs3,cs4}, etc. Moreover, there are hardly any astrophysical sources that are small and dense enough to produce such high-frequency GWs. Therefore, detecting high-frequency GWs would indicate BSM particle physics or the above-mentioned cosmological sources. Thus, on an optimistic note, if the GW detection sensitivity becomes competitive to the DR detection projections in the near future,  unequivocally, a new realm of physics, which is not so well understood in a top-down approach, is ahead of us to explore in a complementary way.   

\section{ Summary}\label{s4}
In this work, we study the possibility of testing non-thermal leptogenesis seeded by inflaton decay with very high-frequency stochastic gravitational waves in the form of gravitons. Inevitable graviton bremsstrahlung is expected when the inflaton decays to the right-handed neutrinos, producing baryon asymmetry via leptogenesis. The negligible influence of thermal scatterings on the production process of lepton asymmetry makes a leptogenesis scenario non-thermal. In the context of the seesaw, it happens when the reheating temperature is smaller than the right-handed neutrino mass scale ($T_{RH} < M$). Neutrino oscillation data on light neutrino masses suggests that a perturbative seesaw Lagrangian corresponds to a $M$ bounded from above.   Therefore, in the non-thermal leptogenesis scenarios, $T_{RH} $ is also bounded. For maximally allowed $T_{RH}$ ($\simeq 8\times 10^{14}$ GeV), we expect a stochastic GW background with frequency $f \gtrsim 10^{11}$ Hz. Although these high-frequency GWs are unlikely to be detected by the proposed high-frequency detectors with their projected sensitivities, for inflaton mass close to the Planck scale, the GWs contribute to sizeable dark radiation within the future sensitivity limit of experiments such as CMB-S4 and CMB-HD (see Fig.\ref{bases}). We also discussed how the future low-energy neutrino experiments that aim to constrain the absolute light neutrino mass scale could complement the high-frequency GW detection to test and constrain non-thermal leptogenesis in specific seesaw models (see Fig.\ref{trhm1}). It would be intriguing to consider such graviton bremsstrahlung as novel probes of non-thermal dark matter and non-thermal co-genesis (simultaneous production of dark matter and baryon asymmetry) formation mechanisms via inflation decay \cite{Samanta:2020gdw,Barman:2021tgt}, which are otherwise notoriously difficult to test in laboratory physics.

\section*{ACKNOWLEDGEMENT}
The work of RS is supported  by the  project International mobility MSCA-IF IV FZU - CZ.02.2.69/0.0/0.0/$20\_079$/0017754 and Czech Science Foundation, GACR, grant number 
20-16531Y. RS also acknowledges European Structural and Investment Fund and the Czech Ministry of Education, Youth and Sports. The work of GW is supported by World
Premier International Research Center Initiative (WPI),
MEXT, Japan. GW was supported by JSPS KAKENHI
Grant Number JP22K14033.
\appendix
\section{The GW energy density spectrum}\label{a1}
\begin{figure} 
\includegraphics[scale=.57]{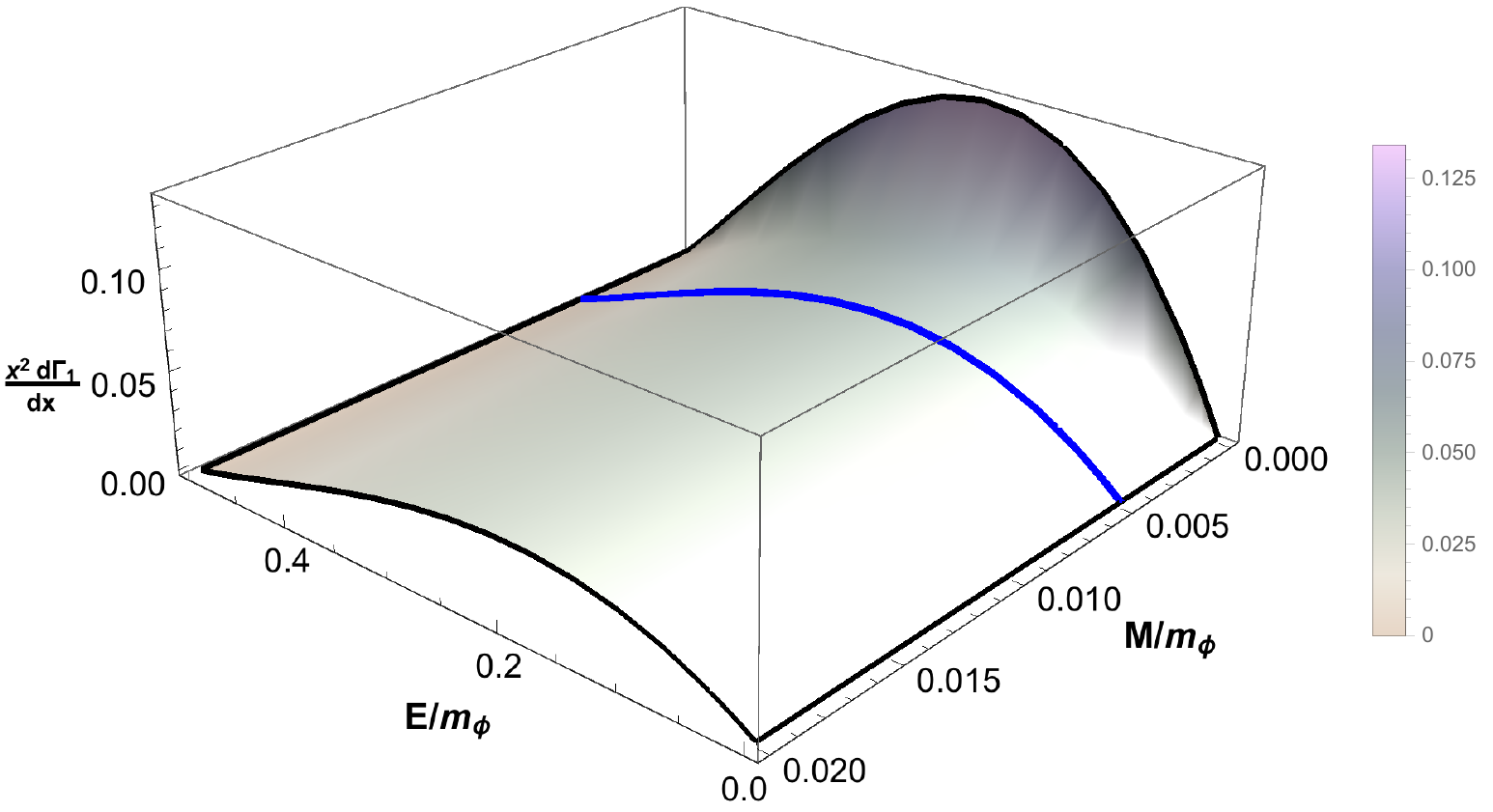}
\caption{ $\frac{x^2 d\Gamma_1}{\Gamma dx }$ vs. $x\equiv E/m_\phi$ vs. $y\equiv M/m_\phi$ plot. The blue arc corresponds to a $\frac{x^2d\Gamma_1}{\Gamma dx}$ that produces the blue curves in Fig.\ref{GWp}. The spectrum peaks around $E\simeq 0.2 m_\phi$.}\label{fig3d}
\end{figure}
Present-day gravitational waves' energy density can be expressed as
\bea
\Omega_{GW}(f)=\frac{f}{\rho_c}\frac{d\rho_g^*}{df}\frac{{a^*}^4}{a_0^4},\label{pgw}
\eea
where $\rho_g^*$ is the energy density of gravitons at a reference temperature $T_*$ and $a$ is the scale factor. Eq.\ref{pgw} can be re-expressed as 
\bea
\frac{f}{\rho_c}\frac{d\rho_g^*}{df}\frac{{a^*}^4}{a_0^4}=\frac{E^2}{\rho_c}\frac{dn_g^*}{dE}\frac{{a^*}^4}{a_0^4}=\frac{E^2}{\rho_c}\frac{n_\phi^*}{\Gamma}\frac{d\Gamma_1}{dE}\frac{{a^*}^4}{a_0^4},
\eea
where we have used $n_g=\frac{\Gamma_1}{\Gamma}n_\phi$. The number density of inflaton can be calculated assuming instantaneous reheating, and it comes out as
\bea
n_\phi=\frac{\rho_R (T_{RH})}{m_\phi(1-\bar{x})},
\eea
where $\rho_R (T_{RH})$ is the radiation energy density at $T=T_{RH}$ and $\bar{x}$ is the energy deposited in the gravitons. Now using $E=x m_\phi$ finally we have
\bea
\frac{f}{\rho_c}\frac{d\rho_g^*}{df}\frac{{a^*}^4}{a_0^4}=\frac{1}{\rho_c}\frac{1}{\Gamma(1-\bar{x})}\frac{x^2d\Gamma_1}{dx}\left[\rho_R (T_{RH})\frac{{a^*}^4}{a_0^4}\right].
\eea

Now considering $T_*=T_{RH}$, the present-day GW energy density is estimated as 
\bea
\Omega_{GW}(f)=\left[\frac{g_\gamma}{g_S (T_{RH})}\right]^{1/3}\frac{\rho_\gamma}{\rho_c}\frac{\Gamma_1/\Gamma}{(1-\bar{x})}\frac{x^2d\Gamma_1}{\Gamma_1 dx},\label{finalgw}
\eea
where assuming no further entropy production after reheating, we use 
\bea
\left[\rho_R (T_{RH})\frac{{a^*}^4}{a_0^4}\right]\simeq \left[\frac{g_\gamma}{g_S (T_{RH})}\right]^{1/3} \rho_\gamma
\eea
with $\rho_\gamma$ being the energy density of photons today. Note that the $x$ in Eq.\ref{finalgw} has to be replaced with $x=2\pi f \frac{T_{RH}}{T_\gamma m_\phi}$ to obtain the present-day $\Omega_{GW}(f)$ vs. $f$ plot as in Fig.\ref{GWp}.

\section{Differential graviton spectrum}\label{b1}
\begin{widetext}
\begin{align}
\frac{d\Gamma_1}{ dx}=  \frac{y_N^2\left(m_\phi/M_{pl}\right)^2}{64\pi^3}&\left[ \frac{2 \left[1+x^2 \left(y^2+2\right)+x \left(-4 y^4+4 y^2-3\right)+y^4-3 y^2\right]}{x} \ln{\left(\frac{1+\alpha}{1-\alpha}\right)} \right. \nonumber \\
{}\;&  \left. + \frac{1+12 x^3+2 x^2 \left(4 y^2-5\right)+4 x y^2+2 y^2}{2x\alpha^{-1}}\right]m_\phi, \label{eq:fermion}
\end{align}
\end{widetext}
where $y=M/m_\phi$ and $\alpha=\sqrt{1-\frac{4y^2}{1-2 x}}$. In Fig.\ref{dgvse}, we show the quantities $\frac{x^2 d\Gamma_1}{\Gamma dx}$ and $\frac{x d\Gamma_1}{\Gamma dx}$ that determine the overall spectral shapes of the GWs and low-energy behaviour of the spectrum respectively. In Fig.\ref{fig3d},  we present a three-dimensional plot of $\frac{x^2 d\Gamma_1}{\Gamma dx}$ with the variables $x$ and $y$. This plot shows irrespective of $y$, the spectrum peaks around $E/m_\phi\simeq 0.2$. Therefore, the present-day peak frequency can be approximated as $f_{\rm peak}\simeq \frac{m_\phi}{10\pi}\frac{T_\gamma}{T_{RH}}$ which in terms of light-neutrino masses can be expressed as 
\bea
f_{\rm peak}^{\rm min}=\frac{m_\phi T_\gamma\left(\tilde{\Phi}\sum_i \sum_km_k |\Omega_{ki}|^2 \right)}{40\pi^2 v^2}.
\eea


\end{document}